\documentclass[graybox]{svmult}

\usepackage{type1cm}        
                            
\usepackage{makeidx}         
\usepackage{graphicx}        
\usepackage{multicol}        
\usepackage[bottom]{footmisc}

\usepackage{hyperref}

\authorrunning{Bulka \textit{et al.}}

\usepackage[dvipsnames]{xcolor}
\usepackage{forest}
\forestset{
  dir tree/.style={
    for tree={
      parent anchor=south west,
      child anchor= west,
      anchor=mid west,
      inner ysep=0pt,
      grow'=0,
      align=center,
      edge path={
        \noexpand\path [draw, \forestoption{edge}] (!u.parent anchor) ++(1em,0) |- (.child anchor)\forestoption{edge label};
      },
      font=\ttfamily,
      if n children=0{}{
        delay={
          prepend={[,phantom, calign with current]}
        }
      },
      fit=band,
      before computing xy={
        l=5em
      }
    },
  }
}

\usepackage{newtxtext}       %
\usepackage[varvw]{newtxmath}       

\makeindex             

\definecolor{brightmaroon}{rgb}{0.76, 0.13, 0.28}

\begin{document}
\title*{Prediction of source nutrients for microorganisms using metabolic networks}
\author{Olivia Bulka \orcidID{0000-0003-1691-6892}, \\ Chabname Ghassemi Nedjad\orcidID{0000-0001-7856-1180},\\ Loïc Paulevé\orcidID{0000-0002-7219-2027}, \\ Sylvain Prigent\orcidID{0000-0001-5146-0347}, and\\ Clémence Frioux\orcidID{0000-0003-2114-0697}}
\institute{Olivia Bulka \at Inria, University of Bordeaux, INRAE, 33400, Talence, France \email{olivia.bulka@inria.fr} 
\and Chabname Ghassemi Nedjad \at Univ. Bordeaux, CNRS, Bordeaux INP, LaBRI, UMR 5800, F-33400 Talence, France AND Inria, University of Bordeaux, INRAE, 33400, Talence, France \email{chabname.ghassemi-nedjad@inria.fr} 
\and Loïc Paulevé \at Univ. Bordeaux, CNRS, Bordeaux INP, LaBRI, UMR 5800, F-33400 Talence, France \email{loic.pauleve@labri.fr}
\and Sylvain Prigent \at Univ. Bordeaux, INRAE, BFP, UMR 1332, F-33140 Villenave d'Ornon, France AND Bordeaux Metabolome, MetaboHUB, INRAE, PHENOME-EMPHASIS, 33140, Villenave d’Ornon, France \email{sylvain.prigent@inrae.fr}
\and Clémence Frioux, \textit{corresponding author} \at Inria, University of Bordeaux, INRAE, 33400, Talence, France \email{clemence.frioux@inria.fr}}
%

\maketitle

\abstract{\newline Metagenomics has lowered the barrier to microbial discovery—enabling the identification of novel microbes without isolation—but cultures remain imperative for the deep study of microbes. Cultivation and isolation of non-model microbes remains a major challenge, despite advances in high-throughput culturomic methods. The quantity of simultaneous experimental variables is constrained by time and resources, but the list can be reduced using computational biology. Given an annotated genome, metabolic modelling can be used to predict source nutrients required for the growth of a microbe, which acts as an initial screen to inform culture and isolation experiments. \newline\indent 
This chapter provides an overview of \textbf{metabolic networks and modelling} and how they can be used to \textbf{predict the nutrient requirements} of a microorganism, followed by a \textbf{sample protocol} using a toy metabolic network, which is then expanded to a genome-scale metabolic network application. These methods can be applied to any metabolic network of interest—which in turn can be created from any genome of interest—and are a starting point for experimental validation of source nutrients required for microorganisms that remain uncultivated to date. } 

\section{Introduction} 
\label{sec:1}

 Microbial populations can be identified ubiquitously in the environment using genome sequencing, but the overwhelming majority remain uncultivated in a lab due to limited knowledge about their natural growth requirements, including their source nutrients \cite{Steen.2019}. Traditional methods for microbial isolation have consisted largely of trial and error by varying rich media recipes, which are biased toward fast-growing nutrient-guzzling ``copiotrophs'', which represent only a small fraction of the breadth of existing microbes \cite{Lagier2015, Hug2016, Salcher2025}. These methods have led to the designation of more than 80\% of identifiable microbes as ``uncultivable''—a phenomenon also referred to as ``the great plate count anomaly'' (\textit{i.e.} there are more microbes in an environmental sample than grow on a plate) \cite{Staley1985}. Technological advances in culturomics have decreased the labour-intensity of traditional methods, but despite these advances, many microbes remain difficult to culture \cite{Lagier2015,Lagier2016,Huang2023}. These yet-uncultivated microbes undoubtedly have far-reaching impacts, from the human gut microbiome to plant health and biogeochemical cycling; their cultivation and isolation are critical bottlenecks in microbiological research. 

\begin{figure}[h!]
\Description{A flow chart showing DNA isolation from a  bacteria of interest and assembly of its genome, followed by reconstruction of a metabolic network, followed by four types of analysis/models to predict source nutrients, resulting in \textit{in vivo} media testing to grow the bacteria of interest (in-lab)}
\includegraphics[width=\textwidth]{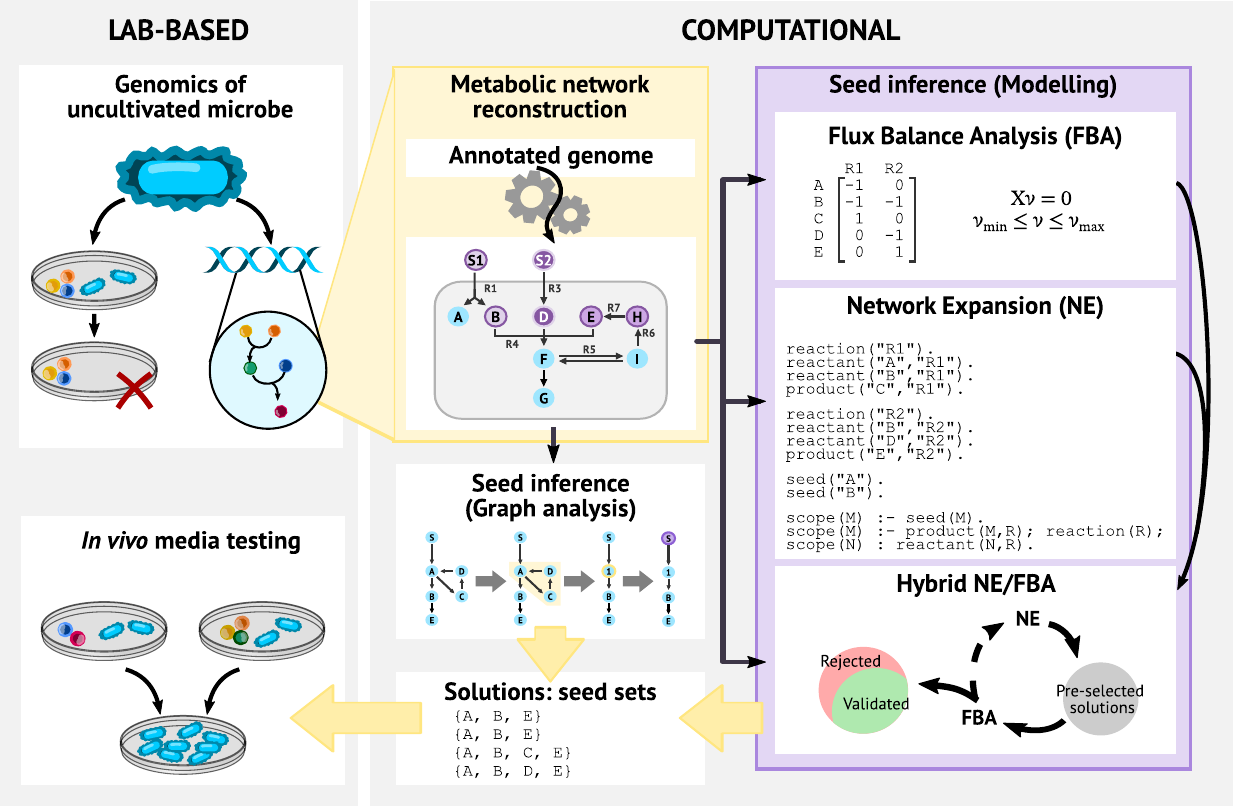}
\caption{Overview of metabolic network-based nutrient inference from genomes for uncultivated microbes. A genome-scale metabolic network needs to be reconstructed from the genome of interest, or obtained from existing knowledge bases. The structure of the network can be used to identify external metabolites (graph analysis), or simulations with several modelling frameworks can be applied to predict source nutrients (seeds). Predicted seeds can then guide experimentation under lab conditions.}
\label{fig:overview}       
\end{figure}

Using [meta]genomics, more information can be gleaned about uncultivated microbes than ever before—whether from lab-enriched microbial communities, endosymbionts cultured intracellularly, or directly from environmental samples \cite{Rinke2013,Lan2022,Stewart2011, Masson2020}. The assembled genomes of such microbes can be used to abstract their metabolic capabilities through the reconstruction of metabolic networks, from which growth of the microbe can be simulated. Such metabolic networks and associated models can also be used to predict the components of growth media for a given microbe. This type of analysis can provide an informed starting point for culturing uncultivated microbes, especially when combined with experimental expertise (Fig.~\ref{fig:overview}).

In this chapter, we introduce the basic concepts of metabolic modelling and provide a narrative protocol or tutorial describing the methods of inferring nutrients, denoted as \textit{seeds}, from a metabolic network. Our objectives are to guide the reader through the main steps, present an overview and comparison of existing models, share reproducible computational experiments, and explain the strengths and limitations of these models—with the ultimate goal of documenting the process of moving from an annotated genome to medium prediction. \\

\begin{svgraybox}
\textbf{Note: }the \textit{italicized} terms are defined in the glossary in the chapter's appendix.
\end{svgraybox} 

\subsection{Modelling metabolic networks}

\begin{figure}[h]
\Description{A simple bacterial cell with seven reaction arrows connecting 12 metabolites to form a network.}
\centering
\sidecaption
\includegraphics[width=0.7\textwidth]{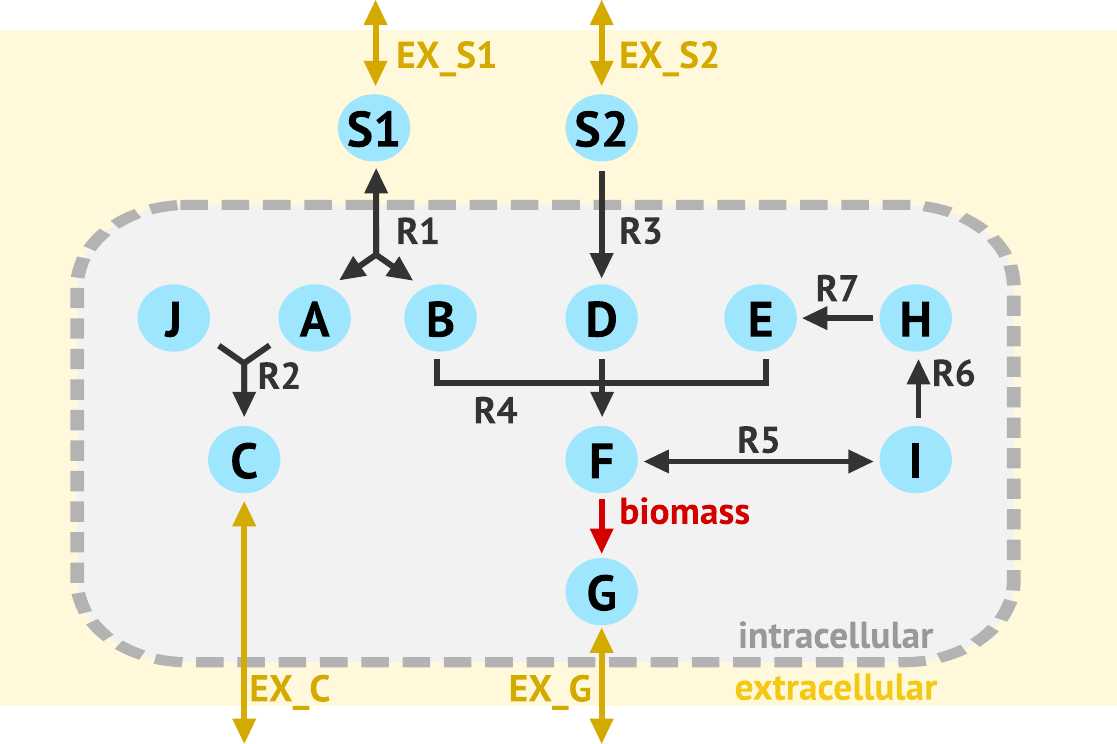}
\caption{Depiction of a toy metabolic network, with metabolites as blue circles and reactions as arrows. The directionality of a reaction is represented by the arrowheads. This network includes two compartments: intracellular (grey) and extracellular (yellow). Exchange reactions illustrating the boundaries of the modelled system are shown in yellow; the biomass reaction is in red.}
\label{fig:network}       
\end{figure}

Metabolic networks represent a web of metabolic reactions that may occur in a cell, where reactions are catalyzed by an enzyme or enzyme complex, converting reactant metabolites to product metabolites. A simple toy model is presented in Figure~\ref{fig:network}, as an example. Reactions can occur in one direction (such as the reaction \verb|R2| in  Fig.~\ref{fig:network}) or both directions (such as \verb|R1| in Fig.~\ref{fig:network}). Metabolic networks can also contain \textit{exchange reactions} to represent transfer of metabolites ``in'' and ``out'' of the network. For example, the exchange reaction \verb|EX_S1| in Figure~\ref{fig:network} imports the metabolite \verb|S1| into the network (or could export it from the network), where it can be consumed by the reaction \verb|R1|. Most metabolic networks also contain a \textit{biomass reaction}, which represents the conversion of all required growth components into cell biomass (DNA, RNA, protein, lipids, etc.) and is a proxy for simulated growth. These are complex reactions involving many metabolites when the network represents a real microbe, but in the toy model, the biomass reaction is represented by a simple reaction highlighted in red in Figure~\ref{fig:network}.

When a metabolic network encompasses all possible metabolic reactions an organism can perform as predicted from its genome annotations, it is referred to as a genome-scale metabolic network (GSMN). GSMNs are used for a number of applications in synthetic biology and industrial biotechnology, such as simulating the impact of a gene deletion on growth or the production of a metabolite of interest. In microbial ecology, GSMNs are often used to simulate a microorganism's growth in different environments, this is commonly referred to as metabolic modelling. GSMNs can also be used to predict which metabolites can be produced and consumed by a microbe in a given environment, or to predict interactions between distinct microbial populations \cite{Cerk.2024}.

Various approaches can be used to study GSMNs. A first strategy can be to rely on \textbf{graph analysis} to explore the connectivity of nodes, paths between compounds of interest, cycles or other centrality metrics \cite{Barabasi2017}.
Alternatively, several modelling approaches enable prediction of a microorganism's activity considering its GSMN, simulated conditions (e.g. medium composition), and an adequate mathematical or computational model. We introduce below two of such modelling approaches, \textbf{network expansion} (NE) and \textbf{flux balance analysis} (FBA), that will be used throughout the chapter. 

\subsubsection{Network expansion}

Network expansion (NE) is a \textbf{qualitative} modelling technique that uses a Boolean approximation to determine the reachability of metabolites in the network given a set of available metabolites, the \textit{seeds} \cite{Romero.2001, Ebenhoh.2004}. In other words, starting from the possible seeds in a microbe's environment or medium, this algorithm determines all possible metabolites that can be reached after flowing through the available metabolic pathways. The resulting collection of compounds reached from the given set of seeds is called the \textit{scope}. 

\begin{figure}[h]
\Description{The simple toy network during three iterations of NE. The first network has two purple metabolites, the second adds two blue metabolites (the products of the reactions where the purple metabolites are reactants), and the third adds three more blue metabolites (the products of the reactions where the blue metabolites were reactants).}
\centering
\includegraphics[width=\textwidth]{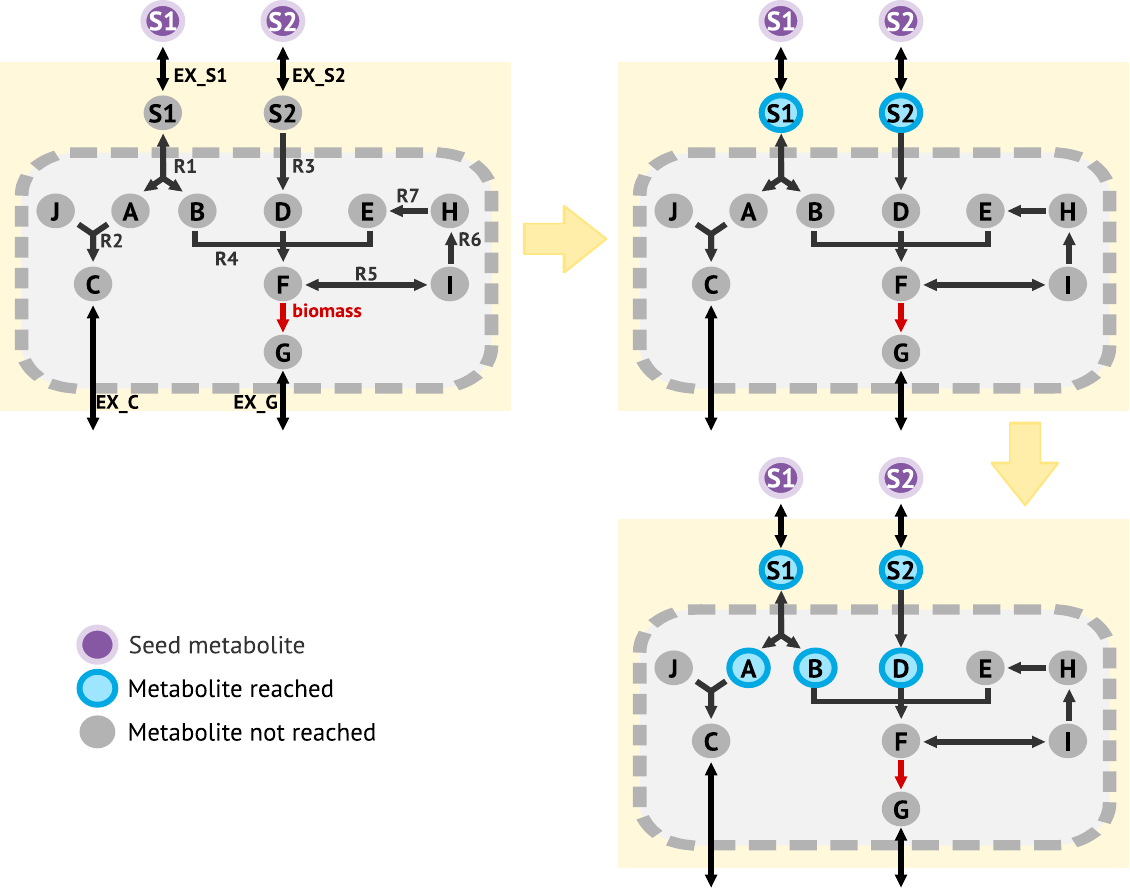}
\caption{Network expansion demonstrated on the toy network, with S1 and S2 as initial seeds (purple). Each iteration expands the reached metabolites (\textit{scope}) by including metabolites newly reachable from those in the previous step (blue). Metabolites that cannot be reached from these seeds remain grey.}
\label{fig:NE}       
\end{figure}

The scope can be computed iteratively by first finding all metabolites that can be produced directly from the provided seeds, then considering those product metabolites in addition to the original seeds as the reactants for the second step. The algorithm continues finding reachable products from the reactant set and moving on to the next step until a fixed point is reached. Figure \ref{fig:NE} illustrates the computation of NE on the toy metabolic network. An important consideration is that every reactant of a reaction must be reached at one step in order for its products to be reached at the next, but the precise stoichiometry is dismissed.  

\subsubsection{Flux balance analysis}
\begin{figure}[h]
\Description{The toy network before and after FBA computation, maximizing the flux in the biomass reaction. After, the reaction arrows are different thicknesses, to represent the quantity of flux through the network.}
\centering
\includegraphics[width=\textwidth]{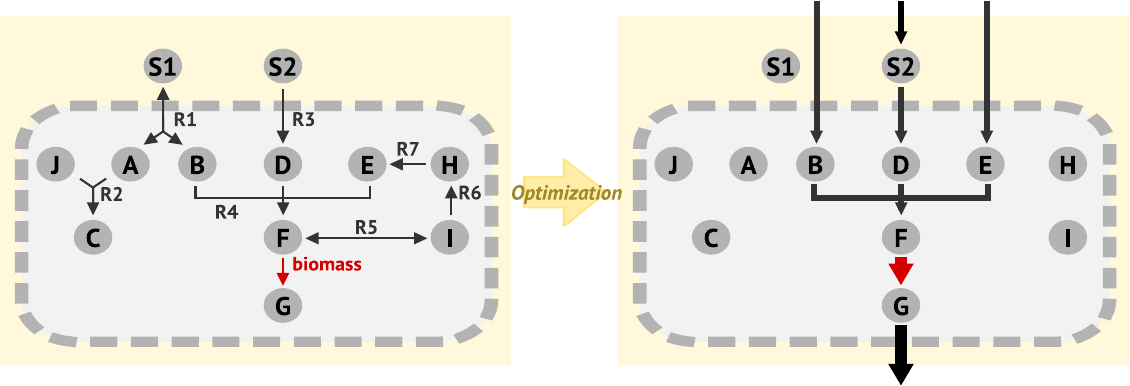}
\caption{Flux balance analysis demonstrated on the toy network. After optimization of the objective function, reaction fluxes are represented by arrows with widths proportional to their flux values. Reactions carrying zero flux are not shown. The \textit{objective function} (biomass reaction) is shown in red.}
\label{fig:fba}       
\end{figure}

Flux balance analysis (FBA) is one of the most commonly used metabolic modelling methods \cite{Orth2010}. Simply, FBA predicts a \textbf{quantitative} map depicting how (and how much of) each metabolite is distributed through a reaction network to produce the most cell biomass possible (a \textit{flux distribution}), while preventing internal accumulation of metabolites. There could be several flux distributions satisfying the constraints, and Figure~\ref{fig:fba} illustrates one example for the toy metabolic network.

FBA uses linear programming to compute a feasible and optimal distribution of reaction \textit{fluxes} (pseudo-reaction rates, usually presented in mmol per gram dry weight per hour) across a metabolic network. It maximizes a user-specified \textit{objective function} (often cellular biomass production, \textit{i.e.} flux through the biomass reaction) subject to thermodynamic and stoichiometric constraints and without accumulation of internal metabolites, \textit{i.e.} assuming the cell is growing at steady-state. For more details on the mathematical constraints of FBA, we refer the interested reader to \cite{Orth2010, Kauffman2003}.  

FBA models include \textit{exchange reactions} to represent the metabolites that can be consumed from the medium (like the previously mentioned \textit{seeds}) or produced by the cell and released. The release of any metabolite specified by an exchange reaction is permitted while satisfying the no-accumulation constraint, as it alleviates a dead end path for that metabolite in the system. Exchange reactions must be explicitly specified by the user.

\subsection{From metabolic networks and models to seed inference} 
\begin{figure}[h]
\Description{A) A network with some purple nodes becoming a network with green nodes, and B) A network with some red nodes becoming a network with green and purple nodes.}
\includegraphics[width=1\textwidth]{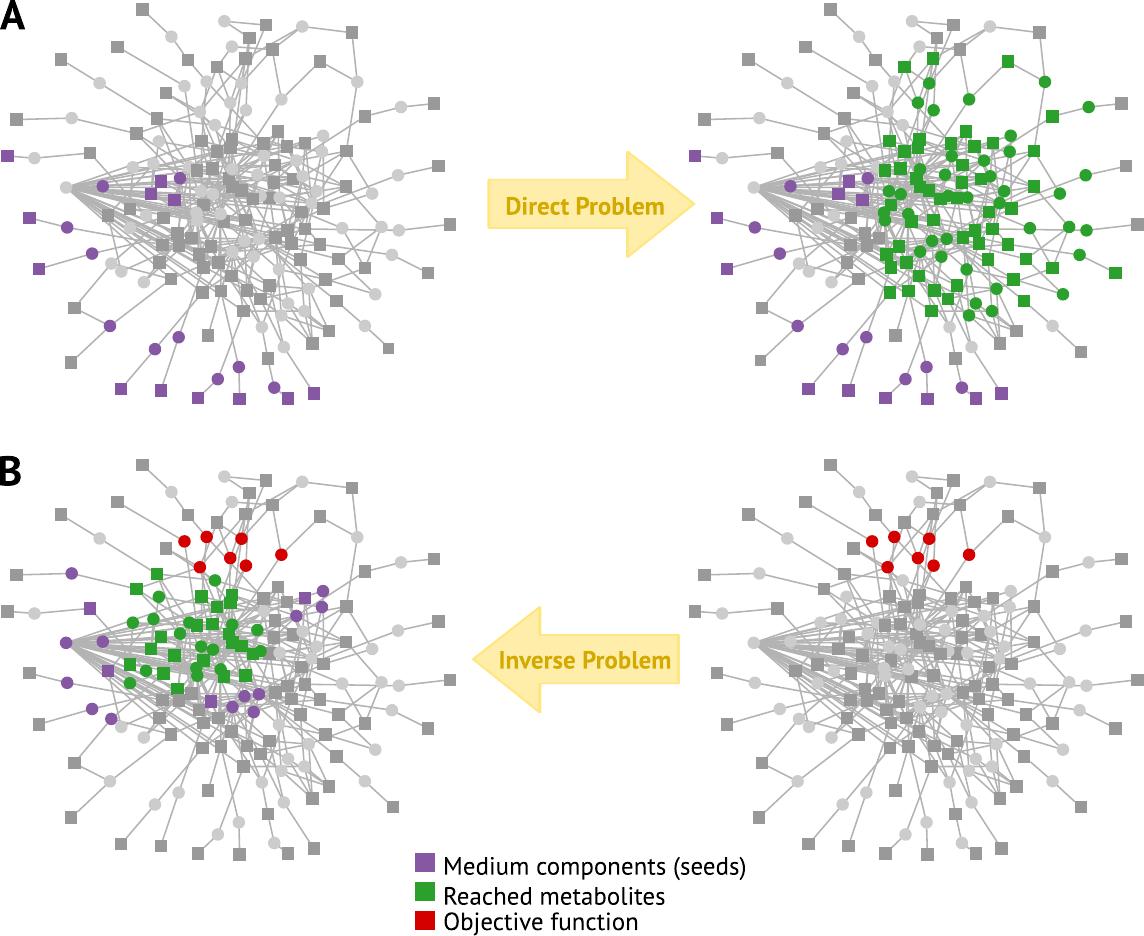}
\caption{Metabolic modelling as a \textbf{A)} direct problem vs. \textbf{B)} an inverse problem (seed inference). Classical metabolic modelling uses known seeds (available nutrients, purple) and a metabolic network (grey) to predict which metabolites can be reached (green), while seed inference does the reverse by applying a known metabolic objective (red) to a network to predict seeds.}
\label{fig:modelvsi}       
\end{figure}

Both NE and FBA modelling typically operate in the forward direction (also called the \textit{direct problem}); they are applied to metabolic networks and environmental conditions (such as metabolites in the medium, \textit{i.e.} seeds) are specified to simulate metabolic activity (\textit{i.e.} growth, metabolite production, etc.). In this chapter, our interest is the prediction of growth conditions, especially growth medium composition, which is referred to as \textit{seed inference}. Seed inference operates in the reverse direction (or the \textit{inverse problem}); it is applied to a metabolic network and certain metabolic information is specified (\textit{i.e.} a behaviour that the model should ensure, such as an objective function), to predict possible seeds in the growth medium or environment (Fig.~\ref{fig:modelvsi}). The same modelling strategies can be applied to this inverse problem, or the network can be analyzed directly without modelling, using topological/graph analysis methods.

\subsection{Seed inference}
 Several methods can be used to infer seed metabolites from an available metabolic network, depending upon the quality of the network reconstruction, available input data, and the biological question.

\subsubsection{Graph-based seed inference}
Topology-based seed inference applies graph theory to the network to explore metabolites by identifying central compounds, paths, metabolite cycles, and highly connected components. 
Tools using this method—like NetSeed \cite{netseed} and PhyloMint \cite{phylomint}—simplify the structure of the network and collapse metabolite cycles into one representative component called a \textit{strongly connected component}. 
Ultimately, a set of seeds that are required to reach all metabolites in the network is identified as the solution. 
This method ignores the stoichiometry of reactions, and cannot target specific reactions or production of specific metabolites, but rather infers seeds based on reaching the full network. In other words, all the metabolic pathways within a network are expected to be attained from the selected seeds, which may not correspond to biological reality. 

\begin{description}[Limitationss]
\item[\textbf{Use Cases}]{exploring the metabolic network of incomplete genomes and poorly-studied organisms to provide qualitative information}
\item[\textbf{Advantages}]{very fast; fewer solutions to parse; fairly intuitive from a network perspective}
\item[\textbf{Limitations}]{can only make full-network inferences; cannot select specific metabolites to target, so the inference cannot be tailored to a targeted phenotype}
\end{description}

\subsubsection{NE-based seed inference}
Like in the forward direction, NE-based seed inference uses Boolean abstraction to signify the presence or absence of metabolites, thus if the reaction metabolites are available, products can be formed regardless of stoichiometry. Though it is typically used to determine which metabolites are reachable from a set of seeds, it can be used to infer seeds from a network, given an objective like biomass production or production of a given metabolite \cite{Romero.2001, Handorf.2008}.

\begin{description}[Limitationss]
\item[\textbf{Use Cases}]{assessing the capabilities of networks derived from incomplete genomes and poorly-studied organisms to provide qualitative information; comparing metabolic capabilities (\textit{scopes}) of different species; modelling growth from an initial state rather than at steady-state}
\item[\textbf{Advantages}]{very fast to execute; robust to draft or poorly balanced models}
\item[\textbf{Limitations}]{lack of accuracy because it does not account for stoichiometry; does not guarantee flux through the objective function}
\end{description}

\subsubsection{FBA-based seed inference}
FBA is typically used to predict flux through a network constrained by stoichiometry and mass balance, assuming that the cell's metabolism is in steady state (\textit{i.e.} no metabolites are accumulating internally) \cite{Orth2010}. It can also be used to solve the reverse problem (for example, by using the \verb|minimal_media| function in COBRApy \cite{Ebrahim2013}), which is to predict which seed metabolites are required in order for flux to reach the objective reaction, while minimizing either the sum of the flux of all exchange reactions or the number of exchanged metabolites.

\begin{description}[Limitationss]
\item[\textbf{Use Cases}]{exploring the metabolism of highly curated GSMNs produced from well-studied organisms with well-annotated genomes; predicting growth yield of organisms in a specific medium}
\item[\textbf{Advantages}]{considers more information than NE, like stoichiometry and thermodynamics; can produce more accurate growth predictions}
\item[\textbf{Limitations}]{incorrect annotation or assumptions can skew results; requires the user to provide a predefined medium; requires manipulation of exchange reactions to obtain best results; getting alternative solutions can be computationally demanding} 
\end{description}

\subsubsection{Hybrid NE/FBA methods}
It is also possible to use hybrid methods that combine NE and FBA to take advantage of the best aspects of each of the two approaches. These methods provide seed solutions that ensure flux through the objective function (like FBA) and also reachability of target metabolites (like NE). There are several hybrid NE/FBA methods made available within tools like Seed2LP \cite{GhassemiNedjad2025}.

\begin{description}[Limitationss]
\item[\textbf{Use Cases}]{fits both curated and non-curated models}
\item[\textbf{Advantages}]{considers more information than NE or FBA alone: leverages NE reachability and FBA's stoichiometry and thermodynamic constrains; can produce more accurate predictions}
\item[\textbf{Limitations}]{slower than NE; high combinatorial space to explore}
\end{description}

\subsection{Goals of this chapter}
This chapter will demonstrate how to predict medium components for a microbe of interest using seed inference from a metabolic network, first with the toy network described in Figure~\ref{fig:network} and then with a genome-scale network as a second example. It will provide the following:

\begin{enumerate}
    \item{Instructions for download and installation of the various computational tools and metabolic networks used in the tutorial}
    \item{An overview of metabolic network selection and network reconstruction from an annotated genome}
    \item{A protocol for application of four existing methods to predict seeds from a metabolic network:
    \begin{enumerate}
        \item{Graph-based/ topological analysis (NetSeed \cite{netseed})} 
        \item{NE (Seed2LP \cite{GhassemiNedjad2025})}
        \item{FBA (COBRApy \cite{Ebrahim2013})}
        \item{Hybrid-NE/FBA (Seed2LP \cite{GhassemiNedjad2025})}
        \end{enumerate}
        }
    \item{A comparison of different strategies for seed inference}
    \end{enumerate}

\section{Methods}
\label{sec:2}
There are many methods one can use to perform seed inference, and this is not an exhaustive review. These protocols aim to provide several options for a user depending on their needs, spanning graph-based analysis and two modelling frameworks, and briefly compare them.

\subsection{Computational requirements}

These analyses can be run on a standard computer on Linux, macOS or Windows, with only one core and 10 Gb RAM. All that is required is a terminal emulator (ex: Terminal, iTerm2, etc.) and a working installation of conda\footnote{\url{https://docs.conda.io/projects/conda/en/latest/user-guide/install}} or Docker\footnote{\url{https://docs.docker.com/engine/install/}}. This protocol assumes the most basic knowledge of terminal use, but several general commands are listed here as a reminder.\\

\begin{warning}{Bash command cheat sheet}
To see which directory you are currently in, enter: \verb|pwd|\\
    \\
    To change directories, enter: \verb|cd [dir_name_here/]|\\ 
    \\
    To list all files in your current directory: \verb|ls|\\
    \\
    To print the contents of a file in your terminal: \verb|cat [path/to/your/file_name.txt]|\\
\end{warning}

\subsection{Software installation}
\label{sec:softinstall}

Software installation can be performed two different ways: 
\textbf{1) using conda} to install the tools in a dedicated environment, or \textbf{2) using Docker} to enhance the reproducibility of the tutorial over time as software versions change. The installation of these tools by each of the two methods is detailed in individual subsections to follow.

In either case, three seed inference tools will be used and compared in this work. \textbf{NetSeedPy}\footnote{\url{https://github.com/cfrioux/NetSeedPy}} performs the topological/ graph-based inference; it is a python re-implementation of NetSeed \cite{netseed} which was originally developed as a downloadable Perl script and a web app but is no longer available. \textbf{Seed2LP}\footnote{\url{https://github.com/bioasp/seed2lp}} implements NE-based and hybrid seed inference \cite{GhassemiNedjad2025}, and \textbf{COBRApy}\footnote{\url{https://cobrapy.readthedocs.io/en/latest/}} includes an FBA-based seed inference method \cite{Ebrahim2013}. 

 A pre-constructed directory structure required to run this tutorial without modification of the commands is also provided for both installation methods. This includes several homemade bash and python scripts, as well some template files for the seed inference tools and the two metabolic networks used in the tutorial (discussed in the Section~\ref{sec:model}). The directory structure and included files are outlined here.

{\small
\begin{forest}
  dir tree
  [seed\_inference\_tutorial/
    [objective/
      [toy-model\_target.txt]
      [iCN718\_target.txt]
    ]
    [sbml/
      [toy-model.xml]
      [iCN718.xml]
    ]
    [scripts/
      [met\_get/
        [met\_get.py]
        [run\_met\_get.sh]
      ]
      [cobra\_seedsearch/
        [cobra\_seedsearch.py]
        [run\_cobra\_seedsearch.sh]
      ]
    ]
    [target/
      [toy-model\_targets.txt]
      [iCN71\_targets.txt]
    ]
  ]
\end{forest}
}

\subsubsection{Direct installation with conda}
\label{sec:conda}
To test if \verb|conda| is already available to use, enter \verb|conda --version| in the terminal. Otherwise, it must be installed according to the documentation\footnote{\url{https://www.anaconda.com/docs/getting-started/miniconda/install}}. Miniconda (Anaconda's smallest distribution) is sufficient for this tutorial. A new conda environment should be created for installation of the seed inference tools. This can be performed by entering the following commands in your terminal emulator of choice.

\begin{programcode}{Bash Code}
    To create and activate a new conda environment with the correct version of python: 
    \begin{verbatim} 
    conda create -n seed-inference python=3.11
    conda activate seed-inference\end{verbatim}
    To install NetSeedPy and Seed2LP: 
    \begin{verbatim}
    pip install git+https://github.com/cfrioux/NetSeedPy.git@main
    pip install seed2lp 
    \end{verbatim}
\textbf{Note:} Seed2LP automatically installs COBRApy, so specific installation is unnecessary.

\end{programcode}

The tutorial directory zip file containing the required scripts and networks is available for download at Recherche Data Gouv\footnote{\url{http://doi.org/10.57745/3Z5L45}}. Once it is downloaded and placed in a desired location, use \verb|cd| in the terminal to move to that location containing the zip archive. Then use the following commands to unzip the archive and prepare for the tutorial. 

\begin{programcode}{Bash Code}
    To unzip and enter the correct directory for the tutorial:
    \begin{verbatim}
    unzip seed_inference_tutorial.zip
    cd seed_inference_tutorial\end{verbatim}
\end{programcode}

\subsubsection{Execution using a Docker image}
\label{sec:docker}
A pre-installed environment for the tutorial is also downloadable as a Docker image. This guarantees that the provided commands will remain repeatable over time as new versions of the software dependencies are released.
To run this tutorial using this image, Docker must first be installed (see its documentation\footnote{\url{https://docs.docker.com/engine/install/}}). The tutorial's Docker image can be found on Docker Hub\footnote{\url{https://hub.docker.com}}, and is also archived at Recherche Data Gouv\footnote{\url{https://doi.org/10.57745/3Z5L45}}. This image includes all tools and tutorial files, rendering the commands in Section~\ref{sec:conda} unnecessary. The docker image can be loaded from either source as follows.

\begin{programcode}{Bash Code}
    From Docker Hub directly:
    \begin{verbatim} 
    docker pull bioasp/seed-inference-tutorial\end{verbatim}
    Or, after downloading the archived docker image from Recherche Data Gouv:
    \begin{verbatim} 
    docker load -i docker-image-seeds.tar 
    \end{verbatim}
    \textbf{Note: }Once the image is loaded, each tutorial command can be executed by prefixing it with \verb|docker run --rm bioasp/seed-inference-tutorial|. For example: 
    \begin{verbatim} 
    seed2lp network sbml/toy-model.xml sbml-norm/ -wf\end{verbatim} 
    would become:
    \begin{verbatim} 
    docker run --rm bioasp/seed-inference-tutorial \
        seed2lp network sbml/toy-model.xml sbml-norm/ -wf\end{verbatim}
\end{programcode}  

\begin{warning}{Important}    
    All the model files are already shipped within the Docker image. To use different input files, their directory can be provided to Docker using the \texttt{-v} option as follows, which exposes the current working directory to Docker:
    \begin{verbatim} 
    docker run -v $PWD:/wd \
        --rm bioasp/seed-inference-tutorial \
        seed2lp  ...\end{verbatim}
\end{warning}

\subsection{Metabolic network selection}
\label{sec:model}
To demonstrate this protocol, we will use two metabolic networks as examples: a toy network (11 reactions, 12 metabolites), and a curated genome-scale network (1015 reactions, 888 metabolites) representing \textit{Acinetobacter baumannii} AYE (iCN718 \cite{Norsigian.2018}). These two models are included in the repository detailed in the \nameref{sec:softinstall} section, but iCN718 can also be found in the BiGG database\footnote{\url{http://bigg.ucsd.edu/static/models/iCN718.xml}} \cite{King.2016}. From BiGG, many curated published GSMNs can be downloaded, which is an efficient and effective method to start modelling with an organism of interest. Fortunately, there are many published networks that can be adapted to a system of interest or used as is \cite{Malik-Sheriff.2019, Karp.2017, King.2016}, but reconstruction of a new network from a given genome is also possible.

\begin{warning}{Important}
\textbf{Note: }Seed2LP is currently only compatible with metabolic networks that use SBML Level 3 Version 2. Please consider this if following the tutorial with a different network.
\end{warning}

\subsubsection{Genome-scale network reconstruction}

As an alternative to using an existing metabolic model, it is possible to create a metabolic network from a genome of interest. GSMNs can be obtained by associating metabolic reactions to the metabolic genes encoded in the genome using gene-protein-reaction (GPR) rules. Instructive reviews exist for undertaking this endeavour \cite{Thiele2010, Gu2019, Ye.2022, Fang.2020, Cerk.2024}.

Many tools can (semi-)automatically create these models from a genome of choice, the most common of which are summarized in Table~\ref{tab:1}. CarveMe \cite{carveme} and GapSeq \cite{gapseq}, for example, are very fast, and do not require manual curation in order to run simulations, but will likely still require curation before these simulations are biologically relevant to the studied organism \cite{Karp.2018}. Before choosing a tool, we suggest referencing a review that compares these tools in greater detail \cite{Gu2019, Mendoza2019}. 

\begin{warning}{Important}
Significant manual curation may be required to obtain a useful GSMN, especially for non-model organisms. Automatic reconstructions can result in incorrect gap-filling, missing gene annotations, absence of unconventional metabolic pathways, inconsistent reactions, incorrect GPRs, missing essential pathways, mass/charge balancing errors, etc. The quality of a metabolic network can be assessed using tools such as BioISO \cite{bioiso} and MEMOTE \cite{memote}, but expert interpretation is required to amend the detected problems. 
\end{warning}

\begin{table}[h]
\caption{Comparison of available metabolic model reconstruction tools \cite{Gu2019,Mendoza2019}. Abbreviations: GUI, graphical user interface.} 
\label{tab:1} 
\begin{tabular}{p{2cm}p{2cm}p{2,4cm}p{1.5cm}p{2.4cm}p{1cm}}
\hline\noalign{\smallskip}
Tool & Language & Source Databases & Gap-filling & Software Type & Citation\\
\noalign{\smallskip}\svhline\noalign{\smallskip}
AuReMe & python & BiGG, MetaCyc & yes & command-line & \cite{aureme} \\
AutoKEGGRec & matlab & KEGG & no & command-line & \cite{autokeggrec} \\%
CarveMe & python & BiGG & yes & command-line & \cite{carveme} \\
CoReCo & python & KEGG & yes & command-line & \cite{coreco} \\%
gapseq & shell, R & KEGG, MetaCyc, ModelSEED, UniProtKB, TCDB & yes & command-line & \cite{gapseq} \\
\textit{merlin} & java & KEGG, MetaCyc, UniProtKB, TCDB & no & GUI & \cite{merlin} \\%
MetaDraft & python & BiGG & no & GUI & \cite{metadraft, metadraft-software} \\
ModelSEED & perl, java & ModelSEED & yes & online interface & \cite{modelseed, modelseed2} \\
Pathway Tools & python, lisp & PGDB, MetaCyc & yes & GUI & \cite{ptools,ptools2} \\%
RAVEN & matlab & KEGG, MetaCyc & yes & command-line & \cite{raven} \\%
\noalign{\smallskip}\hline\noalign{\smallskip}
\end{tabular}
\end{table}

\subsubsection{SBML file structure}

System Biology Markup Language (SBML) \cite{Hucka.2018} is the most widely-used language to share metabolic networks and models, and is the typical input format for metabolic modelling software. These files are computer- and human-readable, and can be opened and inspected with any text editor. SBML files contain all the information about the structure of a metabolic network, including, but not limited to, \textit{compartments}, \textit{metabolites} and \textit{reactions} in a model as well as their metadata.

\verb|listOfCompartments| describes the cellular compartments of the organism, including the cytoplasm and extracellular space, as well as the periplasm or other organism-specific compartments. Reactions are required to explicitly move metabolites between compartments.

\verb|listOfSpecies| includes all metabolites that appear in the network. Each metabolite is described by a unique identifier and usually has a name, a compartment, the associated chemical formula, and other metadata. Metabolite identifiers often begin with \verb|M_| and end with a compartment identifier, such as \verb|_c| for cytoplasm or \verb|_e| for extracellular. This means a metabolite's identifier can change when it moves between compartments.

\verb|listOfReactions| includes biochemical and transport reactions in the network, whose identifiers typically begin with \verb|R_|. Each reaction's description includes its name, the gene(s) associated with the reaction, its reversibility status, and its \textit{bounds} (the maximum and minimum flux allowed) for flux simulation. Each reaction has specified reactants and products, as well as their identifiers and stoichiometry in the reaction.

\begin{warning}{Important}
If a reaction is constrained in the forward direction (lower bound = 0, upper bound $>$ 0), the ``reactants'' in the \verb|listOfReactants| for that reaction are consumed by the reaction and the ``products'' in the \verb|listOfProducts| are produced by the reaction. The opposite occurs if the reaction is constrained in the reverse direction (lower bound $<$ 0, upper bound = 0). In this case, the ``products'' according to the \verb|listOfProducts| are actually consumed by the reaction to produce the ``reactants'' in the \verb|listOfReactants|). When the lower bound is negative and the upper bound is positive, the reaction can operate in both the forward and backward directions.

The designation of reactants vs. products is trivial for FBA, but impacts topological seed inference methods like NetSeed, which do not consider the specified reaction bounds. To NetSeed, the list of reactants is always considered as the reaction substrate—even if the constraints designate the opposite. Network normalization (described in Section~\ref{sec:norm}) can correct this problem by reversing the bounds of the reaction and switching its substrates and products so that the reaction is occurring in the forward direction.
\end{warning}

There are several notable types of reactions that comprise special cases, such as the \textit{biomass reaction} and \textit{exchange reactions}. The biomass reaction is not a biochemical reaction but an abstract representation of the requirements for the organisms to grow. In some types of modelling—like FBA \cite{Orth2010}—this reaction is designated as the objective function for optimization, and if this reaction is activated (\textit{i.e.} holds a positive flux in the simulation), it suggests that the organism can grow in the simulated environmental conditions. Exchange reactions are typically denoted with the prefix \verb|R_EX_|, and have no products. They represent import and/or export of metabolites, as defined by the bounds and reversibility of the reaction (see yellow reactions in Fig. \ref{fig:network}). 
More details about these reactions and other types of ``boundary reactions'' can be found in the COBRApy documentation\footnote{\url{https://cobrapy.readthedocs.io/en/latest/building_model.html}} \cite{Ebrahim2013}.

\section{Protocol applied to a toy network}

This section applies four seed inference methods to the toy metabolic network downloaded with the protocol repository in the \nameref{sec:softinstall} section. The resulting seeds for the toy model are discussed as part of the tutorial, and the application of these methods at genome scale are discussed in Section \ref{sec:3}.

\begin{warning}{Warning}
If this tutorial is paused and revisited, the conda environment must be reactivated when opening a new terminal shell with: \verb|conda activate seed-inference| 

Likewise, always make sure the code is run from the correct directory, which is \verb|seed_inference_tutorial|. The current directory can be checked with \verb|pwd| and can be changed with \verb|cd [dir_name_here/]|.
\end{warning}

\subsection{Network normalization}
\label{sec:norm}
Some seed inference methods include a normalization step that transforms the network in the following ways to ease computation and correct notation that would preclude prediction of relevant results. 

\begin{enumerate}
\item{\textbf{Reversible reactions} are split into two irreversible reactions.} 
\item{Reactions that are \textbf{reversed according to their bounds} (\textit{i.e.} reactions with a negative lower bound and zero as an upper bound) are re-written with a direction reversal, switching products and reactants such that the reaction operates in the forward direction.}
\item{Reactions with \textbf{zero bounds} are removed, as they are expected to be blocked and cannot be used in numerical simulations (like graph-based methods, for example).}
\item{\textbf{Exchange and sink reactions} can be removed to prevent creation of artificial seeds. Keeping the exchange reactions, however, is particularly useful for curated models in which the presence of such reactions represents available metabolites for simulations, \textit{i.e.} seeds.}
\end{enumerate}

\begin{figure}[h]
\Description{This is figure Alt-Text for Figure 1.}
\includegraphics[width=\textwidth]{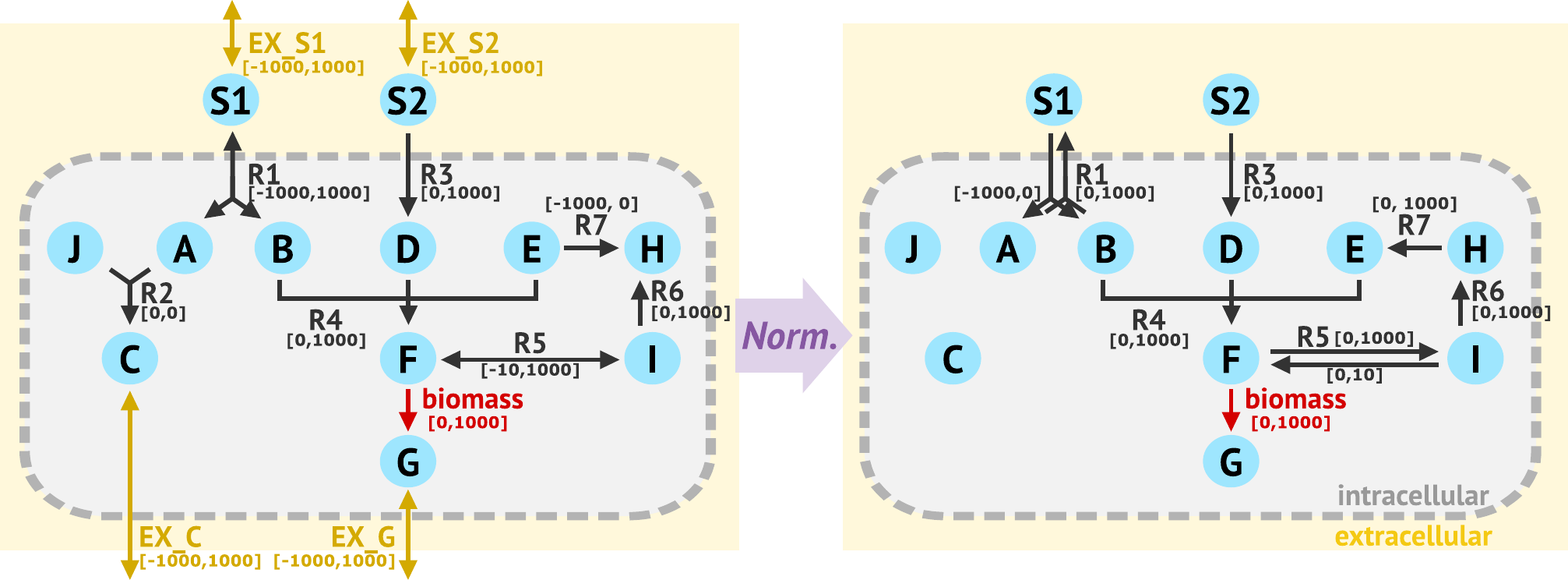}
\caption{Normalization of toy metabolic network, including for steps. \textbf{1)} R1 and R5 were each split into two reactions, one in each direction. \textbf{2)} R7 was reversed so that it operates in the forward direction—its products and reactants were switched and so were its upper and lower bound, such that its notation changes from E$\rightarrow{}$H (reverse) to H$\rightarrow{}$E (forward). \textbf{3)} R2 was removed, as its bounds are both zero. \textbf{4)} All exchange reactions were removed to prevent the presence of artificial seeds.}
\label{fig:net-norm}       
\end{figure}

It is particularly useful to normalize networks before graph-based analysis in order to produce more realistic predictions and ensure comparable results across methods, but it is not an included function of NetSeedPy, for example. 
Therefore, the command to perform normalization using Seed2LP is described in the following box, which should be run before using NetSeedPy directly. The resulting normalized network can be provided to seed inference tools that do not normalize networks (note that Seed2LP will perform normalization by default). 
The transformations resulting from normalization are demonstrated on the toy network in Figure~\ref{fig:net-norm}. 

\clearpage

\begin{programcode}{Bash Code}
    General command structure (fields in square brackets must be specified):
    \begin{verbatim} 
    seed2lp network [path/to/sbml.xml] \
        [path/to/results/directory/] -wf \end{verbatim}
    To normalize the toy model (from the tutorial directory):
    \begin{verbatim} 
    seed2lp network sbml/toy-model.xml sbml-norm/ -wf\end{verbatim}

\end{programcode}

\begin{svgraybox}
\textbf{Result: \textit{Normalization of the toy network using Seed2LP}}\newline

The results should be available within a few seconds, saving a new SBML file at \verb|sbml-norm/toy-model.xml|. The console will print:
\\
\begin{small}
        \begin{verbatim}
        Network name: toy-model
        Objective found for toy-model: R_BIOMASS
        WARNING :
         - R_R2: Deleted.
             Boundaries was: [0.0 ; 0.0]
         - R_R7: Reactants and products switched.
             Boundaries was: [-1000.0 ; 0.0]
        ############################################
        ############################################
                   WRITING SBML FILE
        ############################################
        ############################################
        WARNING : Reaction with tag reversible modified:
        	- R_BIOMASS
        	- R_EX_S1
        	- R_EX_S2
        	- R_EX_C
        	- R_EX_G
        WARNING : Reaction with Reactants and Products exchanged:
        	- R_R7
        WARNING : Import reaction removed:
        	- R_EX_S1
        	- R_EX_S2
        	- R_EX_C
        	- R_EX_G
        File saved at: sbml-norm/toy-model.xml
        \end{verbatim}\end{small}

As the warning above describes, the reaction R2 has zero bounds (\verb|[0,0]|) in the network, meaning no flux is allowed through this reaction. Due to these constraints, the reaction is removed during normalization (Figure~\ref{fig:net-norm}). Additionally, because the bounds of R7 reflect that the reaction is constrained in the reverse direction (\verb|[-1000,0]|), the reactants and products in R7 are reversed (Figure~\ref{fig:net-norm}).  

\end{svgraybox}

\subsection{NetSeed: graph analysis}

The NetSeed python implementation (NetSeedPy) can be run in the terminal in the \verb|seed-inference| conda environment. NetSeedPy must be provided with an SBML file (using the \verb|--sbml| flag), which will be parsed to extract metabolites and reactions before graph simplification and seed inference. The output format is specified using \verb|--format enumeration|, which prints one solution per line as comma-separated metabolite identifiers, limiting the number of solutions to 100 using the \verb|--max| flag.

\begin{programcode}{Bash Code}
\begin{nopagebreak}
Command structure:
\begin{verbatim} 
netseedpy --sbml [path/to/sbml.xml] \
    --format enumeration --max [No. solutions]
\end{verbatim} 
\end{nopagebreak}
For the toy model:
\begin{verbatim} 
netseedpy --sbml sbml-norm/toy-model.xml \
    --format enumeration --max 100\end{verbatim}

\end{programcode}

\begin{figure}[h]
\sidecaption[t]
\Description{The toy network, with predicted seeds highlighted in purple.}
\includegraphics[width=7.8cm]{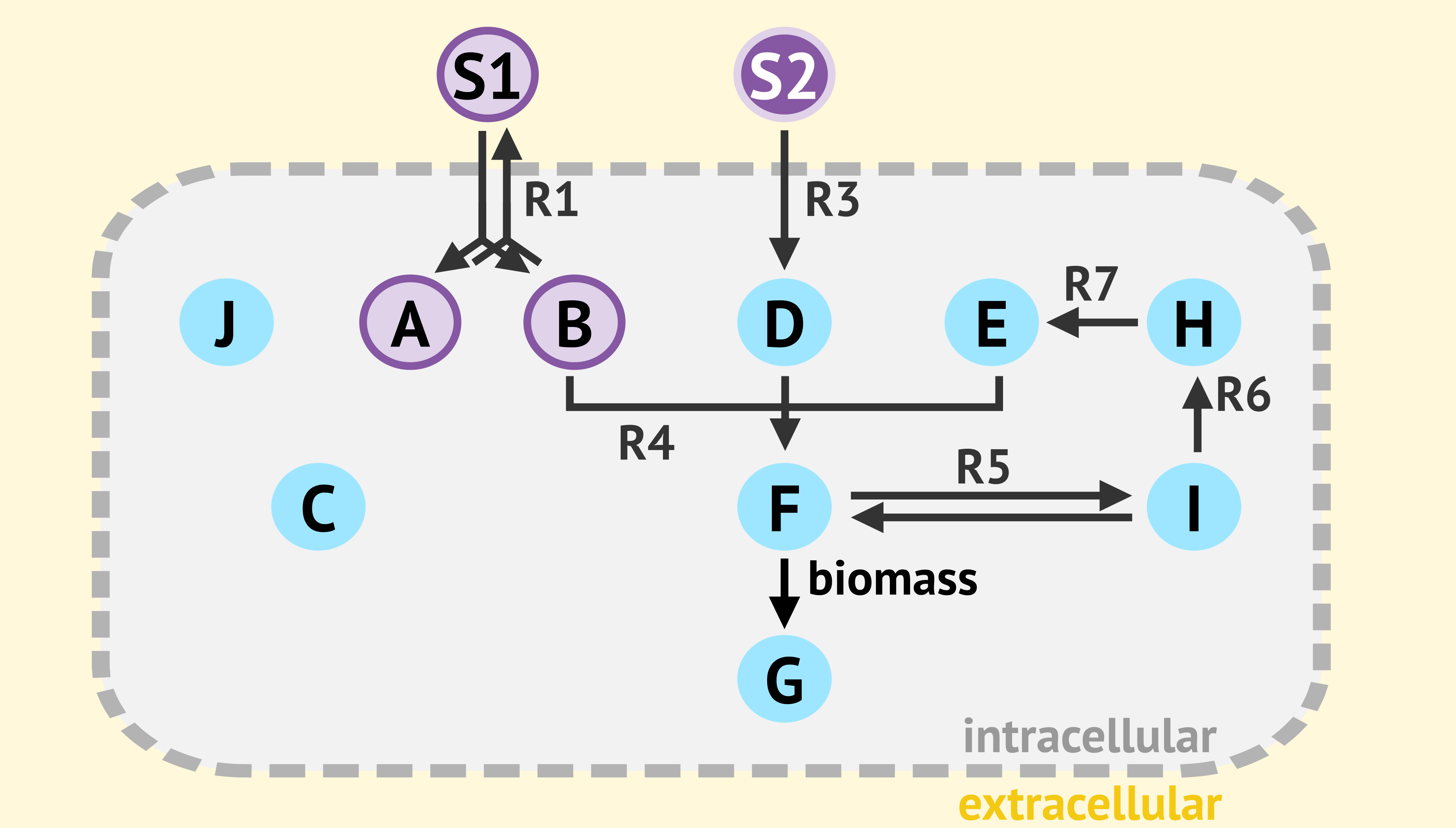}
\caption{Predicted seed sets (three solutions) for the toy network using graph analysis via NetSeedPy. Possible seeds are coloured purple, with the same shade of purple indicating those seeds which are interchangeable according to the solution sets. In this case, one of (S1 or A or B) belongs to each set, whereas (S2) belongs to all solution sets. }
\label{fig:toy-netseed}       
\end{figure}

\begin{svgraybox}
\textbf{Result: \textit{Graph analysis of the toy network using NetSeedPy}}\newline

The results should be available within seconds, showing there are three possible solutions with two seeds each (Fig.~\ref{fig:toy-netseed}). 

\begin{small}\begin{verbatim}   

                Solution_1 -- M_A_c, M_S2_e
                Solution_2 -- M_B_c, M_S2_e
                Solution_3 -- M_S1_e, M_S2_e

\end{verbatim}\end{small}

Each metabolite is listed by its identifier, which includes the prefix \verb|M_| and the suffix \verb|_e| or \verb|_c|, depending on whether it is extracellular or in the cytosol, respectively. These prefixes and suffixes are removed from visualizations for clarity.
\end{svgraybox}

\subsection{Seed2LP} 

For this tutorial, Seed2LP will be used for NE-based and hybrid-NE/FBA seed inference \cite{GhassemiNedjad2025}. For both, we will use \verb|target| mode, which automatically detects the metabolic objective in an SBML model, if its coefficient (\verb|fbc:coefficient|) is 1. For information on how to customize the metabolic objective, see the Seed2LP documentation\footnote{\url{https://github.com/bioasp/seed2lp}}. 

Throughout this tutorial, we will use the submode ``subset minimization'' using \verb|-m subsetmin|, which predicts only minimal subsets (ex: if the set (A,B) is a solution, we know that any superset containing (A,B) is also a solution—so sets such as (A,B,C) that contain (A,B) as a subset are not minimal, and would not be listed as solutions). 

For additional info and troubleshooting, see the Seed2LP documentation.

\subsubsection{Network Expansion}
 To run the simplest model of network expansion, we will select the reasoning method using the \verb|-so reasoning| option, and use the \verb|--accumulation| flag to allow the accumulation of internal metabolites by the reasoning method during enumeration, which stays true to the original definition of NE \cite{Ebenhoh.2004}. For each command, an SBML file containing the metabolic network must be provided, as well as the name of an output folder in which results will be written.
 
\begin{programcode}{Bash Code}
Command structure:
\begin{verbatim} 
seed2lp target [path/to/sbml.xml] [path/to/results/directory/] \
    -so reasoning -m subsetmin \
    -nbs [No. solutions] -tl [time limit in mins] \
    --accumulation
\end{verbatim}
\begin{samepage}
For the toy model:
\begin{verbatim} 
seed2lp target sbml/toy-model.xml results/s2lp/toy-model/ \
    -so reasoning -m subsetmin -nbs 100 -tl 10 \
    --accumulation
\end{verbatim}
\end{samepage}
\end{programcode}

\begin{figure}[h]
\Description{Two depictions of the toy network, each with different metabolite nodes coloured purple.}
\includegraphics[width=\textwidth]{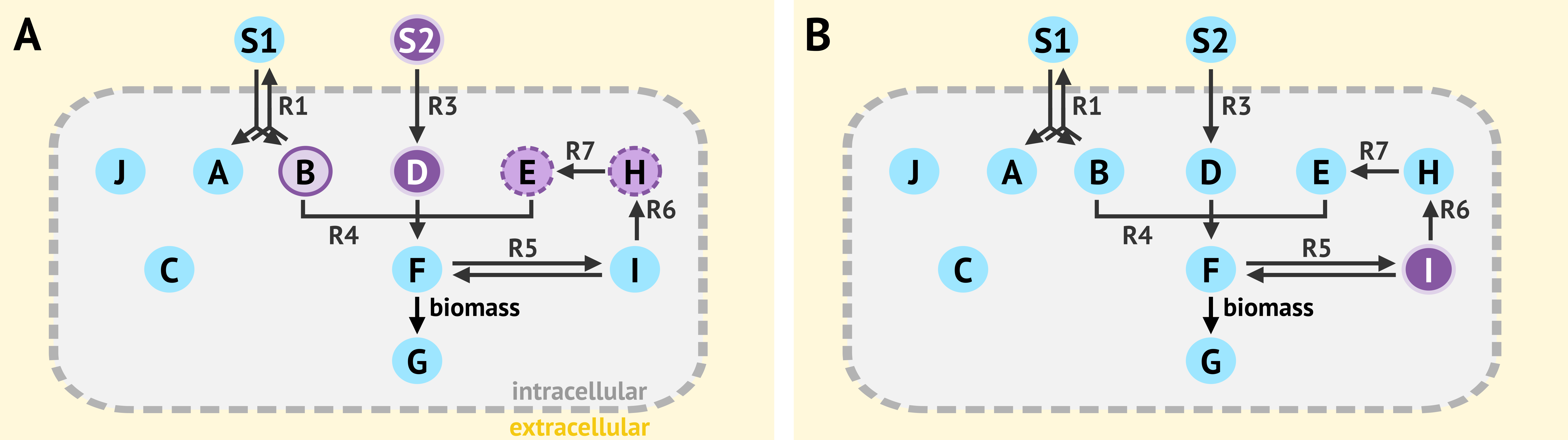}
\caption{Predicted seeds for the toy network Seed2LP for Hybrid-NE/FBA. Seeds are coloured purple, and alternative seeds are the same shade. \textbf{A)} shows four solutions with three seeds each: (B) and (S2 or D) and (E or H), and \textbf{B)} shows one solution (I).}
\label{fig:toy-hybrid}       
\end{figure}

\begin{svgraybox}
\textbf{Result: \textit{NE on the toy network using Seed2LP}}\newline

The results should be available within several seconds, showing nine possible solutions. One solution has one metabolite (I), and eight solutions have three seeds each: (B or S1) and (H or E) and (S2 or D) (Fig.~\ref{fig:toy-hybrid}). Seed2LP should print the following sets to the console:

            \begin{small}\begin{verbatim}   
            Mode : TARGET
            Option: TARGETS ARE FORBIDDEN SEEDS
            ACCUMULATION: Authorized
            Time limit: 10.0 minutes
            Solution number limit: 100
            ____________________________________________
            ____________________________________________
                        Sub Mode: SUBSET MINIMAL
            ____________________________________________
            ____________________________________________
            ················ Classic mode ···············
            ~~~~~~~~~~~~~~~~ Enumeration ~~~~~~~~~~~~~~~
            SOLVING...
            Answer: 1 (1 seeds)
                M_I_c
            Answer: 2 (3 seeds)
                M_B_c, M_H_c, M_S2_e
            Answer: 3 (3 seeds)
                M_B_c, M_D_c, M_H_c
            Answer: 4 (3 seeds)
                M_B_c, M_D_c, M_E_c
            Answer: 5 (3 seeds)
                M_B_c, M_E_c, M_S2_e
            Answer: 6 (3 seeds)
                M_D_c, M_E_c, M_S1_e
            Answer: 7 (3 seeds)
                M_E_c, M_S1_e, M_S2_e
            Answer: 8 (3 seeds)
                M_D_c, M_H_c, M_S1_e
            Answer: 9 (3 seeds)
                M_H_c, M_S1_e, M_S2_e
            
            TIME  DATA EXTRACTION : 0.003s
            TIME TOTAL SEED SEARCH: 0.176s
            TIME       TOTAL      : 0.18s
            \end{verbatim}\end{small}
            \textbf{Note: }The default behaviour of Seed2LP includes cytosolic compounds (metabolites whose IDs end with ``\verb|_c|'') as possible seeds.
\end{svgraybox}

\begin{figure}[h]
\Description{Two depictions of the toy network, each with different metabolite nodes coloured purple.}
\includegraphics[width=\textwidth]{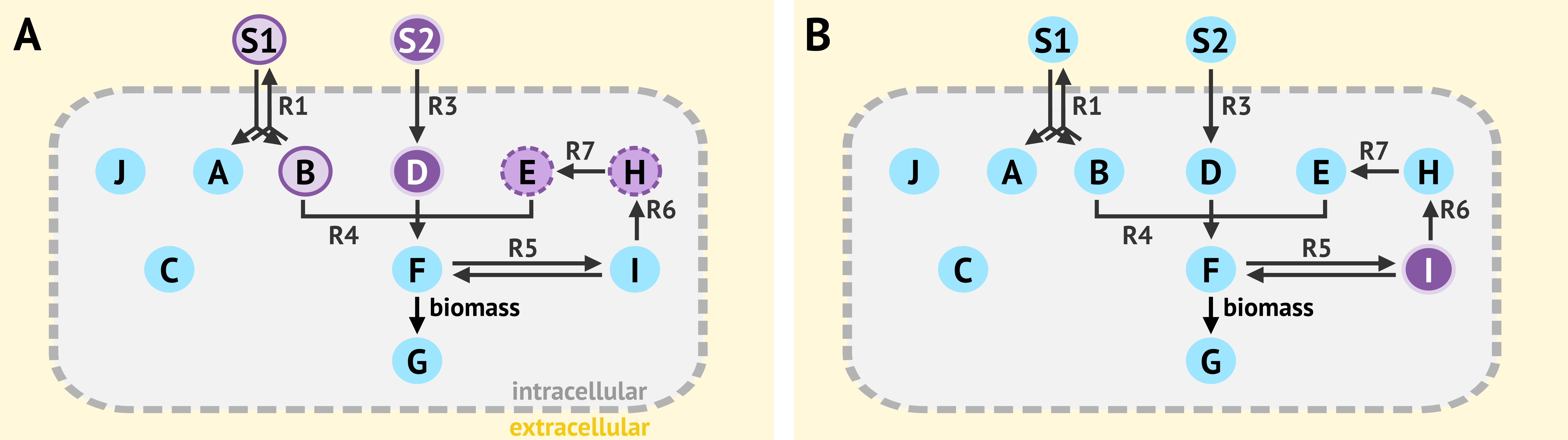}
\caption{Predicted seeds for the toy network using Seed2LP for NE. Seeds are coloured purple, and alternative seeds are the same shade. \textbf{A)} shows eight solutions with three seeds each: (either S1 or B) and (S2 or D) and (E or H), and \textbf{B)} shows one solution (I).}
\label{fig:toy-NE}       
\end{figure}

\subsubsection{NE/FBA Hybrid}

There are several hybrid seed inference methods included in Seed2LP that can be explored, but this tutorial will only consider the ``Guess-and-check with diversity'' method (\emph{Hybrid-GC$_{div}$}), following the outcomes of the original publication \cite{GhassemiNedjad2025}. This mode is provided to the tool with the \verb|-so guess_check_div| option. For the sake of comparison with the reasoning mode, classical NE with the \verb|--accumulation| flag will be considered here again. 

\begin{programcode}{Bash Code}
Command structure:
\begin{verbatim} 
seed2lp target [path/to/sbml.xml] [results/directory/] \
    -so guess_check_div -m subsetmin\
    -nbs [No. solutions] -tl [time limit in mins] \
    --accumulation
\end{verbatim}
For the toy model: 
\begin{verbatim} 
seed2lp target sbml/toy-model.xml results/s2lp/toy-model/ \
    -so guess_check_div -m subsetmin \
    -nbs 100 -tl 10 \
    --accumulation\end{verbatim}
\end{programcode}

\begin{svgraybox}
\textbf{Result: \textit{Hybrid-NE/FBA with the toy network using Seed2LP}}\newline

The results for the toy model should be available within several seconds, showing one solution with one seed (I) (Fig.~\ref{fig:toy-NE}B), and four solutions with three seeds each (Fig.~\ref{fig:toy-NE}A): B and (H or E) and (S2 or D). Seed2LP should print to the console:
            \begin{small}\begin{verbatim}
            Mode : TARGET
            Option: TARGETS ARE FORBIDDEN SEEDS
            ACCUMULATION: Authorized
            Time limit: 10.0 minutes
            Solution number limit: 100
            ____________________________________________
            ____________________________________________
                        Sub Mode: SUBSET MINIMAL
            ____________________________________________
            ____________________________________________
            ····· Guess-Check with diversity mode ······
            ~~~~~~~~~~~~~~~~ Enumeration ~~~~~~~~~~~~~~~
            Answer: 1 (1 seeds)
            M_I_c
            Answer: 2 (3 seeds)
            M_B_c, M_D_c, M_E_c
            Answer: 3 (3 seeds)
            M_B_c, M_E_c, M_S2_e
            Answer: 4 (3 seeds)
            M_B_c, M_D_c, M_H_c
            Answer: 5 (3 seeds)
            M_B_c, M_H_c, M_S2_e
            Rejected solution during process: 4
            
            TIME  DATA EXTRACTION : 0.003s
            TIME TOTAL SEED SEARCH: 3.002s
            TIME       TOTAL      : 3.005s
            \end{verbatim}\end{small}

 The number of rejected solutions are also printed to the console (in this case there are 4). This describes the number of seed sets deemed solutions by NE that are not feasible by FBA, therefore ultimately rejected.
All these solutions satisfy both the NE reachability of the objective reaction's reactants, and a positive flux in the FBA objective reaction, provided that exchange reactions for the corresponding seeds are created.
\end{svgraybox}

\subsection{COBRApy: steady-state–based inference}
\label{sec:cobra}

COBRApy includes a function called \verb|minimal_medium|, which can predict the fewest number of metabolites required to produce biomass in the model \cite{Ebrahim2013}. The COBRApy implementation cannot select any compound for which no exchange reaction is defined (\textit{i.e.} internal metabolites cannot be seeds), which limits the solution space to those metabolites explicitly selected by the model creator. To overcome this limitation and explore the possible seeds most widely, especially for non-curated models, exchange reactions for every internal metabolite in the model can be created before application of COBRApy's FBA-based seed inference method. 

Creation of exchange reactions for each internal metabolite, followed by FBA-based seed inference can be executed by a single command to invoke a homemade script stored in \verb|scripts/cobra_seedsearch/| in the downloadable directory. 
For a given model named ``\verb|MODEL.xml|'', this script requires a target file (\verb|target/MODEL_targets.txt|) and an objective file (\verb|objective/MODEL_target.txt|), which are each a text file containing the reaction ID of one objective reaction (or a list of target metabolites with one ID per line, in the case of the target file). The target file is to ensure that the reactants of the objective reaction are forbidden as seeds (for example \verb|M_F_c|, in the case of the toy model), and the objective file is to set the objective function for simulations using COBRApy. These files are provided automatically in the tutorial directory structure.

\begin{programcode}{Bash Code}
The general command structure:

\begin{verbatim}
python ./scripts/cobra_seedsearch/cobra_seedsearch.py \
    [path/to/sbml.xml]  [path/to/target.txt]  \
    [path/to/objective.txt] [results/directory/] \
    [No. simulations]
\end{verbatim}

For the tutorial data, run the following (all the files are already created):
\begin{verbatim}
python ./scripts/cobra_seedsearch/cobra_seedsearch.py \
    ./sbml/toy-model.xml  ./target/toy-model_targets.txt \
    ./objective/toy-model_target.txt ./results/cobrapy 1\end{verbatim}
\end{programcode}

\begin{figure}[h]
\sidecaption[t]
\Description{The toy network with only one purple metabolite.}
\includegraphics[width=7.8 cm]{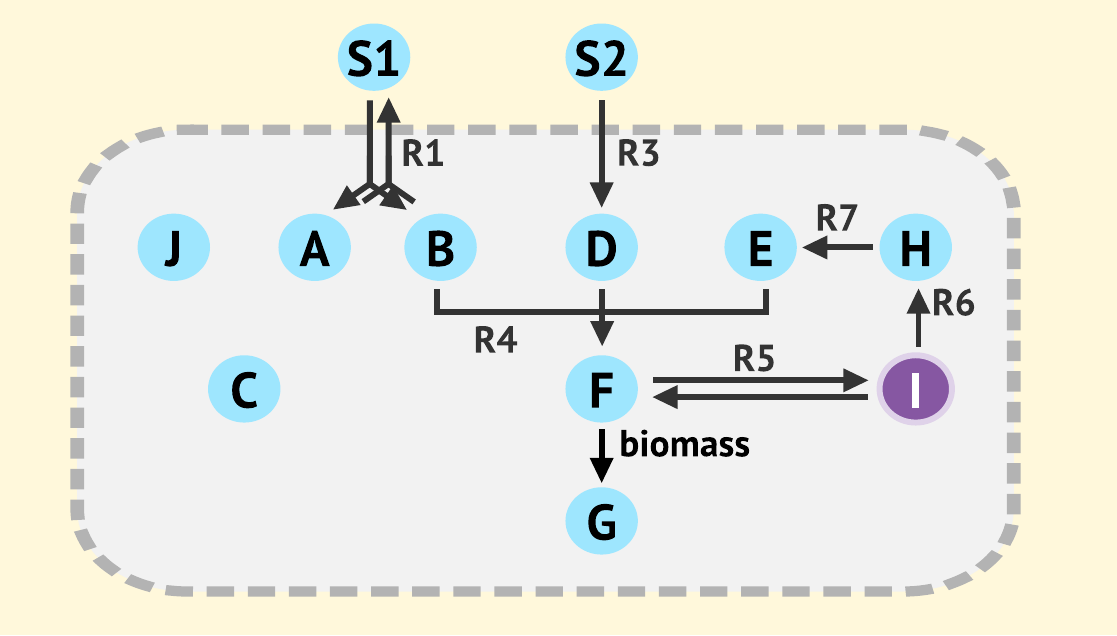}
\caption{The only predicted seed (I, purple) for the toy network using COBRApy via Seed2LP.}
\label{fig:toy-fba}       
\end{figure}

\begin{svgraybox}
\textbf{Result:\textit{ FBA-based seed inference on the toy network using COBRApy}}\newline

The results should be available within minutes, producing a result file called \verb|toy-model_results.json| in a subfolder of \verb|results/cobrapy/|. The JSON file (copied below) contains one solution with one seed (I) (Figure~\ref{fig:toy-fba}).
\begin{small}\begin{verbatim}

        	"OPTIONS": {
        		"REACTION": "All metabolite as exchange reaction",
        		"ACCUMULATION": "NA",
        		"FLUX": "has flux (= 0,1)"
        	},
        	"NETWORK": {
        		"NAME": "toy-model",
        		"SEARCH_MODE": "Cobrapy",
        		"OBJECTIVE": [
        			"R_BIOMASS\n"
        		],
        		"SOLVE": "Cobrapy"
        	},
        	"RESULTS": {
        		"Cobrapy": {
        			"ENUMERATION": {
        				"solutions": {
        					"model_1": [
        						"size",
        						1,
        						"Set of seeds",
        						[
        							"M_I_e"
        						]
        					]
        				},
        				"time": 0.078
        			}
\end{verbatim}\end{small}
As stated in the \verb|FLUX| section of the output file, the biomass reaction holds a positive flux (0.1 \textit{mmol}$\cdot$\textit{g dry weight}\textsuperscript{-1}$\cdot$\textit{h}\textsuperscript{-1}) after optimization with FBA.
\end{svgraybox}

\subsection{Interpretation of Results (Toy Network)}
\begin{figure}[p]
\Description{Three toy networks with seed metabolites in purple. Solutions are listed on the side.}
\includegraphics[width=\textwidth]{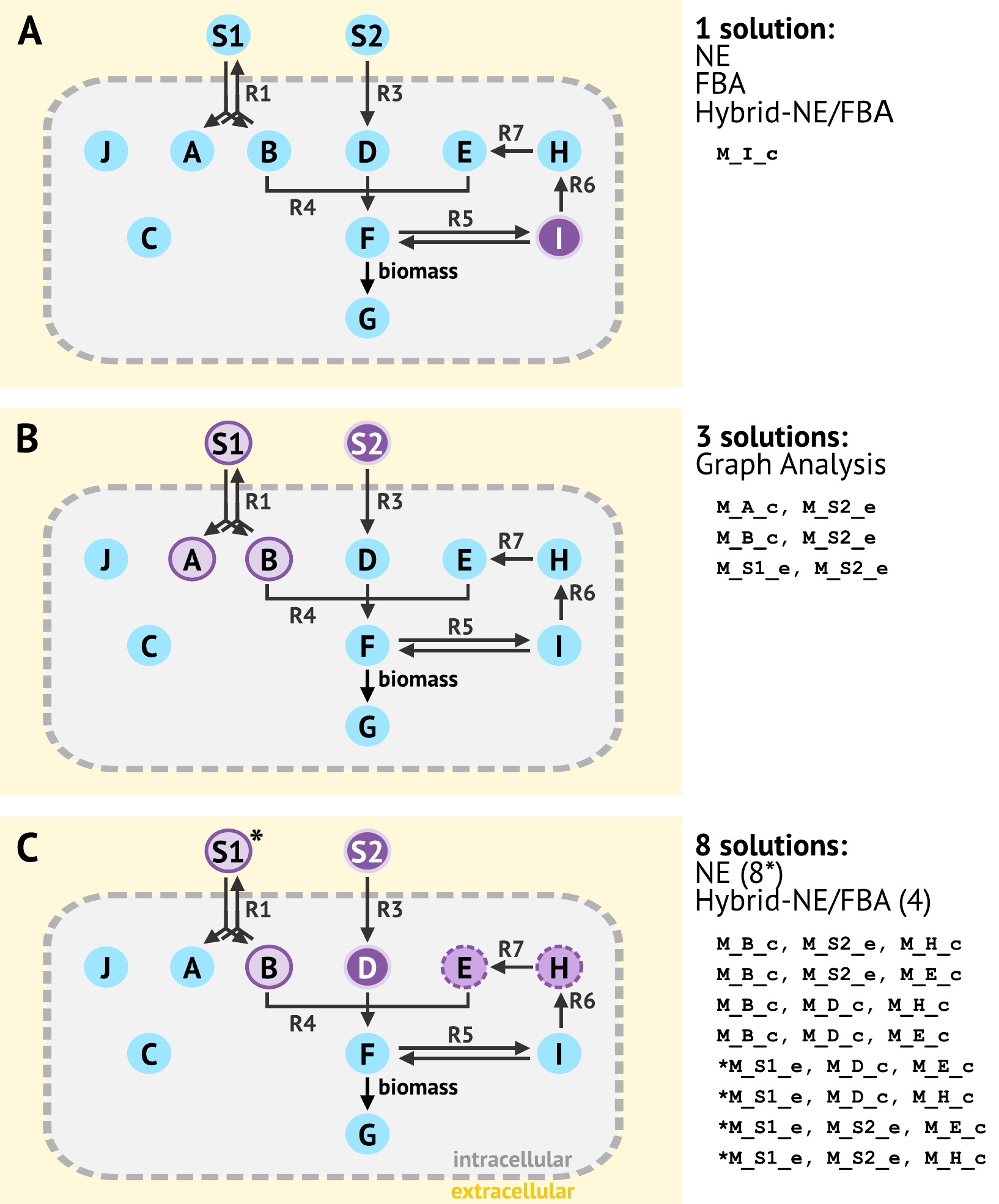}
\caption{Comparison of all predicted seed solutions for the toy network. Possible seeds are coloured purple, with the same shade of purple indicating those seeds which are interchangeable according to the solution sets. \textbf{A)} one one-seed solution. \textbf{B)} three two-seed solutions. \textbf{C)} eight three-seed solutions are predicted, four of which (those including S1, starred) of which were only predicted by NE.}
\label{fig:toy-all}       
\end{figure}

 All solutions predicted in this tutorial and their modelling frameworks are summarized in Figure~\ref{fig:toy-all}. Across the four prediction methods, twelve different solutions are suggested—one one-seed solution (predicted by NE, FBA, and hybrid-NE/FBA; Figure~\ref{fig:toy-all}A), three two-seed solutions (predicted by graph analysis; Figure~\ref{fig:toy-all}B), and eight three-seed solutions (four of which are predicted by NE and hybrid-NE/FBA, and four of which—those involving S1, starred—are only predicted by NE; Figure~\ref{fig:toy-all}C). 

\begin{figure}[h]
\sidecaption[t]
\Description{A bar chart.}
\includegraphics[width=7.8cm]{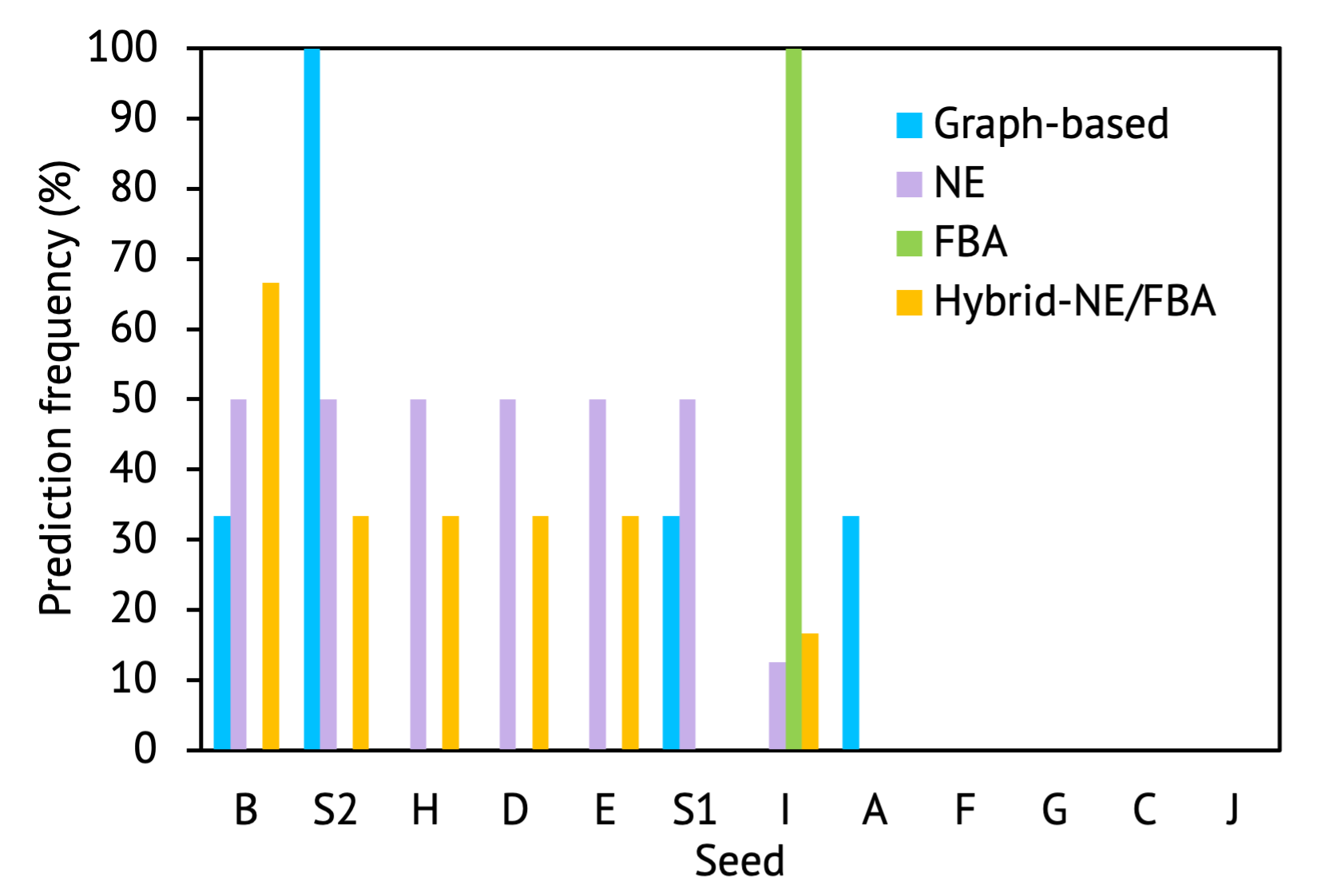}
\caption{The frequency of each seed in the solution sets predicted for the toy network using graph analysis (blue), NE (purple), FBA (green), or Hybrid-NE/FBA (yellow) methods. Frequency is presented as a percent of the total number of solutions per method.}
\label{fig:met_get_toy}       
\end{figure}

The frequency of the predicted metabolites across all predicted seed sets can be compared, which can approximate relevance of the seeds (Figure~\ref{fig:met_get_toy}). For a genome-scale model, this data may be useful for determining which experimental leads to follow first.

\section{Results: application at genome-scale}
\label{sec:3}

\subsection{Seed inference on a genome-scale model}

The same methods can be applied to a genome-scale model, for example \textit{Acinetobacter baumannii} AYE (iCN718; 888 metabolites, 1015 reactions). This strain is of medical interest, as it is a multi-drug resistant pathogen and biotechnological pathogen thanks to its exploitable metabolic and enzymatic capabilities \cite{Norsigian.2018}. The commands to perform seed inference using the same methods as above just require a change in the input SBML file, and the name of the desired results folder (see below).

\begin{programcode}{Bash Code}
    Graph analysis using NetSeedPy, after network normalization using Seed2LP:
    \begin{verbatim} 
    seed2lp network sbml/iCN718.xml sbml-norm/ -wf
    
    netseedpy --sbml sbml-norm/iCN718.xml \
        --format enumeration --max 100
    \end{verbatim}
    \begin{samepage}
    NE using Seed2LP:
    \begin{verbatim} 
    seed2lp target sbml/iCN718.xml results/s2lp/iCN718/ \
        -so reasoning -m subsetmin -nbs 100 -tl 10 \
        --accumulation
    \end{verbatim}
    \end{samepage}
    FBA-based using COBRApy:
    \begin{verbatim} 
    python ./scripts/cobra_seedsearch/cobra_seedsearch.py \
        sbml/iCN718.xml  ./target/iCN718_targets.txt \
        ./objective/iCN718_target.txt ./results/cobrapy 1
    \end{verbatim}
    Hybrid-NE/FBA using Seed2LP:
    \begin{verbatim} 
    seed2lp target sbml/iCN718.xml results/s2lp/iCN718/ \
        -so guess_check_div -m subsetmin \
        -nbs 100 -tl 10 --accumulation
    \end{verbatim}

\end{programcode}

\subsection{Comparison of seed inference methods}

The four tested methods varied in terms of their number of predicted solutions, time taken, and the size of predicted seed sets (Table~\ref{tab:2}). The metabolic objective was the production of biomass precursors (NE mode), positive flux in the biomass reaction (FBA mode), or both (Hybrid-NE/FBA). The methods using the simplest computational models like NetSeedpy and the NE Seed2LP method were the fastest, allowing the NE method to produce the most solutions given the time limit. With an increased time limit and access to a computing cluster or server, the number of solutions found using the more computationally intense methods (like Hybrid NE/FBA) can be increased.

\begin{table}[h]
\caption{Comparison between seed inference methods on a genome-scale model (iCN718), when limited to 100 solutions and 10 minutes, including which constraint limited the number of solutions produced by each method, if relevant.} 
\label{tab:2} 
    \begin{tabular}{p{4cm}p{1.5cm}p{2cm}p{2cm}p{2cm}}
    \hline\noalign{\smallskip}
         & Runtime (s) & No. Solutions & Seed Set Size & Limitations \\
        \noalign{\smallskip}\svhline\noalign{\smallskip}
        Graph-based (NetSeedPy) & 2 & 1 & 141 & \\
        NE (Seed2LP) & 6 & 100 & 8-15 & max solutions\\
        FBA (COBRApy via Seed2LP) & 45 & 1 & 1 &  \\
        Hybrid-NE/FBA (Seed2LP) &  600 & 10 & 10-15 & time\\
    \noalign{\smallskip}\hline\noalign{\smallskip}
    \end{tabular}
\end{table}

\begin{figure}[h]
\Description{This is figure Alt-Text for Figure 1.}
\includegraphics[width=\textwidth]{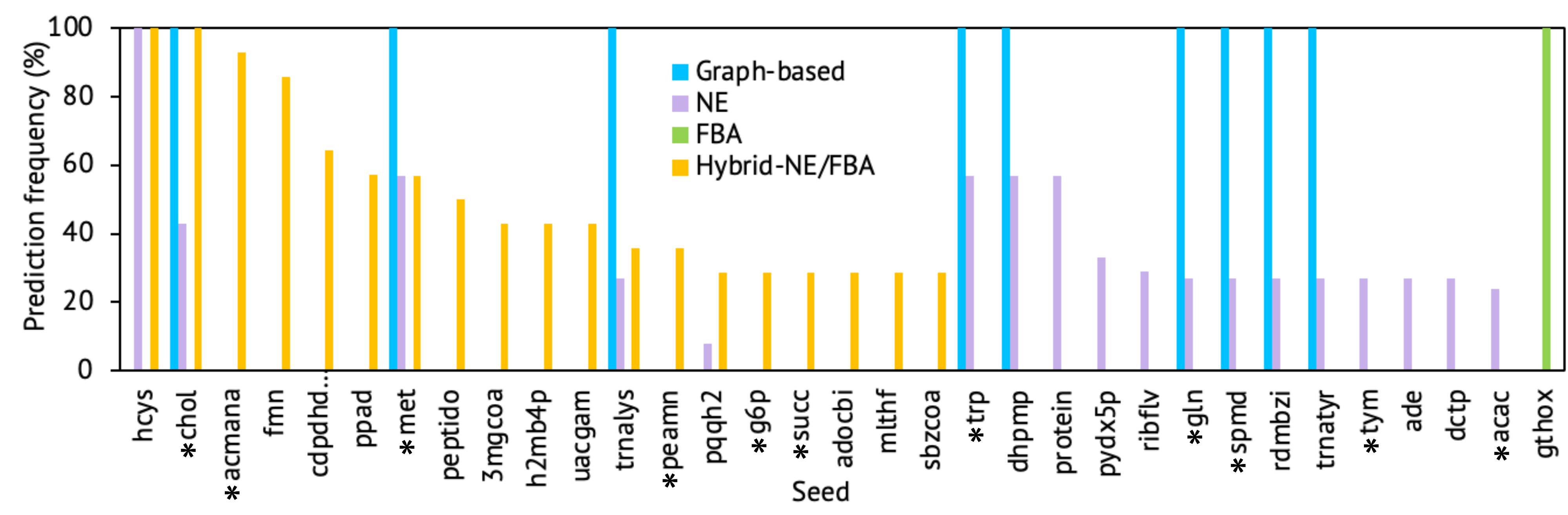}
\caption{The frequency of each seed in the solution sets predicted for iCN718 network using graph analysis (blue), NE (purple), FBA (green), or Hybrid-NE/FBA (yellow) methods. Frequency is presented as a percent of the total number of solutions per method, and only seeds predicted in $>$25\% of modelling solutions are shown. Starred metabolites have an associated exchange reaction in the original SBML file. }
\label{fig:met_get_iCN}       
\end{figure}

The most frequently predicted seeds from each of the seed inference methods were extracted and compared in Figure~\ref{fig:met_get_iCN}. Though frequency of prediction does not necessarily correlate with biological relevance, observing the most common seeds can provide information about the metabolic network, which facilitates a starting point for deeper analysis of the solutions and \textit{in vivo} culturing tests. 

Of the top 33 metabolites, only 11 had an associated exchange reaction specified in the metabolic network (Figure~\ref{fig:met_get_iCN}), which correspond to the simulated medium in the published model. The 22 other top predicted seeds only existed as cytosolic metabolites, without a transport pathway into or out of the cell. If using the \verb|minimal_medium| function in COBRApy directly, these cytosolic metabolites could not be predicted as possible seeds.  For most microorganisms, the extent of metabolite import and export potential is relatively understudied, as very few transporter proteins have been biochemically characterized. Testing multiple seed inference methods and permitting the exchange of many metabolites can help overcome our limited knowledge in this area, or at least reduce the impact of knowledge biases established due to over-representation of certain bacteria in microbiological studies.

\section{Discussion}

Seed inference from metabolic networks is an efficient and practical way to predict possible source nutrients as a first step towards designing defined media for a microorganism. Defined media design is a process that plagues many fields in microbiology—from the human gut to marine microbiology to agriculture \cite{Huang2023, Lagier2016, Masson2020, Eilers2000, Gaspari2020}. Almost every field of microbiology includes certain microbes that cannot yet be cultured, whether due to interdependence on a microbial community, unknown growth requirements, or reliance on an endosymbiotic host \cite{Masson2020}. This demand, combined with the increased accessibility of genomic data, positions seed inference as an invaluable method for microbiologists in many fields to adopt.

\subsection{Model limitations}

Though metabolic modelling and seed inference are excellent tools for informed hypothesis generation and explanation of observed phenomena, they have many limitations to consider. These methods cannot predict many key metrics necessary for cell culture, such as concentrations of medium components, pH, and temperature. Likewise, biological information like enzyme or reaction kinetics \cite{Link2014,Islam2021} and gene regulation \cite{Li2023} are not taken into account using these methods. Increasingly complex types of metabolic modelling are required for these interests \cite{ Carlson2024, King2015}.

Predicted medium components are only useful if they are usable. Seed inference tools may predict metabolites that are too expensive or difficult to work with. Such metabolites can (and perhaps should) be forbidden from the predicted seed sets. Likewise, if expert knowledge insists upon a certain metabolite in the seed solutions, it can be forced in the solution set when using tools such as Seed2LP, for instance. This is useful if the microbe of interest is already cultivable, but perhaps relies on a very complex medium. In this case, seed inference can be performed with the goal of predicting simpler media in which the microbe can grow. Sometimes, however, even the experts have little idea of the possible metabolites imported or exported by the microbe, and in this case a more exploratory approach can be taken and considering all metabolites, including internal ones, as potential seeds can be a relevant strategy.

Metabolic modelling efforts are most impactful when followed by careful experimentation to confirm or qualify predictions. Innovations in culturomics can be used to test many predicted media recipes simultaneously in addition to those parameters that cannot easily be tested using this type of model, like temperature, pH, and oxygen availability \cite{Huang2023,Lagier2016}. Validation in the lab is key to truly understanding the metabolism of any microbe.

\subsection{Applications for microbial communities}

Growth of an isolated microbe is not always the goal, especially since interspecies interactions in more complex microbial communities are pertinent to health and environmental applications \cite{Pacheco.2019m3q, Pande.2017, Ma.2025}. Inferring media for growth of multispecies microbiota may also be of interest. In this case, one must consider not only which nutrients are required for each microbe in the community to grow, but also which nutrients can be transferred between microbes through cross-feeding and which may be subject to competition across population. The same seed inference principles can be applied and extended to small communities where these interspecies exchanges are also accounted for \cite{GhassemiNedjad2025}. 
As the size of the community increases, more computational memory and time is needed to perform simulations, so use of computing cluster or server may be necessary (ex: for a community of 4 GSMNs, the memory use increased to 50 GB in \cite{GhassemiNedjad2025}). 

\subsection{Importance of media design}
The ability to study microbes using computational methods like metagenomics and metabolic modelling does not diminish the importance of the ability to grow axenic cultures, which is critical for microbial characterization. Cultures enable characterization of cell physiology, which is prerequisite to de-orphan uncharacterized genes and analyze microbial community structure and function. Current libraries of medium recipes were designed with the biases of previously-cultured microbes—namely human-associated microbes or other fast-growing microbes who thrive in rich media \cite{Richards.2014}. New strategies are needed for isolation and growth of those microbes who thrive under different conditions \cite{Salcher2025, Staley1985, Eilers2000, Steen.2019}.
When used in tandem with experimental data and expert knowledge to facilitate validation, seed inference and metabolic modelling is a valuable strategy for reducing the impact of the culturability bottleneck.

\paragraph{\textbf{Data Availability}}
The zip file containing the tutorial's directory structure, metabolic networks, and scripts is archived at
Recherche Data Gouv: \href{https://doi.org/10.57745/3Z5L45}{doi.org/10.57745/3Z5L45}. 
The tutorial's Docker image can also be found in the same Recherche Data Gouv repository, but also on
Github: \href{https://github.com/bioasp/docker-seed-inference-tutorial}{bioasp/docker-seed-inference-tutorial}
and Docker Hub: \href{https://hub.docker.com/r/bioasp/seed-inference-tutorial}{bioasp/seed-inference-tutorial}.

\paragraph{\textbf{Contributions}} 

\textbf{O.B.}: 
conceptualization (equal); 
data curation (equal);
software (equal); 
validation (lead);
visualization (lead); 
writing – original draft preparation (lead);  
writing – review and editing (equal).
\textbf{C.G.N.}: 
software (equal); 
validation (supporting);
visualization (supporting); 
writing – original draft preparation (supporting); 
writing – review and editing (equal).
\textbf{L.P.}: 
data curation (equal);
software (equal); 
validation (supporting);
writing – review and editing (equal).
\textbf{S.P.}: 
conceptualization (equal);
validation (supporting);
writing – review and editing (equal).
\textbf{C.F.}: 
conceptualization (lead); 
software (equal); 
validation (supporting);
writing – original draft preparation (supporting);  
writing – review and editing (equal)\\

\begin{acknowledgement}
This work received support from the French government, managed by the National Research Agency (Agence Nationale de la Recherche) under the France 2030 initiative, reference ANR-24-RRII-0003, and operated through the INRAE EXPLOR'AE programme. This work was also supported by the MetaboHUB infrastructure (MetaboHUB ANR-11-INBS-0010) and ANR France 2030 PEPR Systemes Alimentaires, Microbiome et Santé CULTISSIMO ANR-24-PESA-0002.

\end{acknowledgement}
 \section*{Appendix}
 \addcontentsline{toc}{section}{Appendix}

\begin{description}[scope]
\item[\textbf{biomass reaction}]{a reaction that represents the conversion of all required growth components into cell biomass as a proxy for simulated growth; often the objective function in FBA modelling}
\item[\textbf{bounds}]{constraints stating the minimum (lower bound) and maximum (upper bound) flux allowed through a reaction during optimization in flux balance analysis; typically represented as (-1000,1000), for example}
\item[\textbf{compartments}]{a representation of a defined cellular region, such as cystosol or the periplasm; the extracellular space is also a compartment included in most metabolic models}
\item[\textbf{direct problem}]{calculation or prediction of the behaviour of a given system; in metabolic modelling, a model applied to a metabolic network and provided information about seeds in order to simulate metabolic activity (\textit{i.e.} growth, metabolite production, etc.)}
\item[\textbf{exchange reactions}]{representations of metabolites transfer ``in'' and/or ``out'' of the system represented by a metabolic network, including consumption from the medium or production and release of a metabolite by the cell; typically denoted with the prefix \verb|R_EX_|}
\item[\textbf{flux}]{a vector representing the amount of metabolites flowing through a given reaction (usually presented in mmol metabolite per gram dry cell weight per hour)}
\item[\textbf{flux distribution}]{a set of fluxes representing all reactions in a network that satisfy the constraints of flux balance analysis; one specific FBA solution}
\item[\textbf{inverse problem}]{calculation or determination or identification of a system from knowledge of its behaviour; the reverse of typical metabolic modelling, wherein a model is applied to a metabolic network and provided information about metabolic activity (\textit{i.e.} growth, etc.) in order to predict seeds}
\item[\textbf{metabolite}]{any compound that can interact with a reaction in a metabolic network, including organic metabolites, ions, gasses, or even water; sometimes more complex molecules like coenzymes, cofactors, and lipids are included as metabolites, depending on the purpose of the model}
\item[\textbf{objective function}]{the user-defined ``metabolic goal'' of the simulated microbe; often cellular biomass production (\textit{i.e.} flux through the biomass reaction), but can also be the production or consumption of a metabolite of interest}
\item[\textbf{reaction}]{the conversion of one or more metabolites into one or more different metabolites OR the transport of one or more metabolites between compartments; some reactions are those which are catalyzed by an enzyme \textit{\textit{in vivo}}, but others may be spontaneous}
\item[\textbf{scope}]{a set of all the metabolites that are reachable from a given set of seeds, usually referred to in network expansion modelling}
\item[\textbf{seed inference}]{prediction of source nutrients (seeds) for growth of a microbe given a metabolic network}
\item[\textbf{seeds}]{the essential source nutrients required for growth of a microbe, or for simulation of growth/metabolite production in a metabolic network}
\item[\textbf{strongly connected component}]{(or SCC) a group of metabolites in a network that can each be reached from any other metabolite within that group (a cycle); an SCC can be compressed during topology/graph-based analysis to simplify the network}
\end{description}

\bibliographystyle{spmpsci}
\bibliography{biblio}
\end{document}